\documentclass[twocolumn]{aastex6}
\usepackage{amsmath}
\usepackage{multirow}
\usepackage{graphicx}

\begin{document}


\title{Super-Earth LHS3844\MakeLowercase{b} is tidally locked}

\author{Xintong Lyu, Daniel D.B. Koll\altaffilmark{1}}
\affiliation{Peking University \\
Beijing, China}

\altaffiltext{1}{dkoll@pku.edu.cn}

\author{Nicolas B. Cowan}
\affiliation{McGill University \\
Montreal, Canada}

\author{Renyu Hu}
\affiliation{Jet Propulsion Laboratory, California Institute of Technology \\
Pasadena, CA, USA}

\author{Laura Kreidberg}
\affiliation{Max-Planck Institute for Astronomy \\
Heidelberg, Germany}

\author{Brian E.J. Rose}
\affiliation{University at Albany \\
Albany, NY, USA}

% 250 word limit

\begin{abstract}
Short period exoplanets on circular orbits are thought to be tidally locked into synchronous rotation. If tidally locked, these planets must possess permanent day- and nightsides, with extreme irradiation on the dayside and none on the nightside. However, so far the tidal locking hypothesis for exoplanets is supported by little to no empirical evidence. Previous work showed that the super-Earth LHS 3844b likely has no atmosphere, which makes it ideal for constraining the planet's rotation. Here we revisit the Spitzer phase curve of LHS 3844b with a thermal model of an atmosphere-less planet and analyze the impact of non-synchronous rotation, eccentricity, tidal dissipation, and surface composition. Based on the lack of observed strong tidal heating we rule out rapid non-synchronous rotation (including a Mercury-like 3:2 spin-orbit resonance) and constrain the planet’s eccentricity to less than $\sim0.001$ (more circular than Io’s orbit). In addition, LHS 3844b's phase curve implies that the planet either still experiences weak tidal heating via a small-but-nonzero eccentricity  (requiring an undetected orbital companion), or that its surface has been darkened by space weathering; of these two scenarios we consider space weathering more likely. Our results thus support the hypothesis that short period rocky exoplanets are tidally locked, and further show that space weathering can significantly modify the surfaces of atmosphere-less exoplanets.
\end{abstract}


\section{Introduction}
\label{sec:introduction}

Exoplanets on short period orbits are widely believed to be tidally locked to their host stars. Tidal interactions between planet and star tend to align the planet's rotation period about its own axis with its orbital period. The final outcome of this process is that the planet's orbital and rotation periods become equal, a state called 1:1 tidal locking or synchronous rotation  \citep{dole1964,kasting1993habitable,barnes2017}. In such a state the planet always faces the star with the same hemisphere, creating a permanent day- and nightside. The resulting instellation asymmetry is expected to significantly impact the climates of short period exoplanets, including their thermal emission maps, their atmospheric dynamics, and their potential habitability \citep[e.g.,][]{joshi1997,showman2002,yang2014low}.

Not all short period exoplanets need to be synchronously rotating, however.
A planet's spin evolution is complicated by many factors including the planet's internal structure, the presence of an atmosphere or ocean, and the potential influence of companion planets \citep{barnes2017}.
As a historical example, Mercury was long believed to be synchronously rotating based on early tidal theory and observations of its surface features \citep{lowell1898,sheehan2011figure}. Despite these expectations, radar observations decades later revealed that Mercury is in a 3:2 spin-orbit resonance instead \citep{colombo1965rotation}.

Similarly, tidal models predict a wide range of possible spin states for short period exoplanets. These states include Mercury-like spin-orbit resonances \citep{makarov2012,makarov2012a}; chaotic rotation \citep{correia2019}; and complex non-synchronous states for planets with atmospheres \citep{leconte2015asynchronous} or oceans \citep{auclair-desrotour2019}, with similar non-synchronous states plausible for planets that have internal magma oceans or other fluid layers. Whether any particular exoplanet is in such a state is uncertain. For example, Mercury-like spin-orbit resonances are not possible if a planet's eccentricity is zero. 
However, tidal models permit stable spin-orbit resonances at eccentricities as low as $e=0.001-0.01$ \citep{correia2014,correia2019}, which is smaller than the precision to which one can typically constrain exoplanet eccentricities from radial velocity measurements or transit light curves \citep{shen2008,kane2012,vaneylen2015,sagear2023}.
%As a result it is difficult to rule out non-synchronous rotation on the basis of tidal models alone.
As a result it is difficult to assert that short period exoplanets are synchronously rotating solely based on tidal models.

Ideally, one would thus constrain exoplanet rotation rates empirically. For hot Jupiters at least the bulk atmospheric rotation can be inferred via Doppler broadening of absorption lines or via phase curve measurements \citep{snellen2014,kempton2014,rauscher2014,brogi2016}. In practice, however, it is difficult to obtain precise constraints from such observations. For example, the Doppler broadening of HD 189733b is compatible with synchronous rotation, but to within 1$\sigma$ the planet could also be in 3:2 or 2:1 rotation \citep{brogi2016}. Similarly, the unusual westward hot spots found in the phase curves of hot Jupiters Kepler-7b, Kepler-12b, Kepler-41b, and CoRoT-2b might be due to non-synchronous rotation, but they could alternatively also be caused by clouds or magnetic drag \citep{rauscher2014,hu2015a,shporer2015,dang2018}.

In contrast to hot Jupiters, it should be relatively straightforward to infer the rotation of atmosphere-less exoplanets (bare rocks for short).
The thermal phase curve of a bare rock primarily depends on the planet's rotation, its orbit, the surface's thermal inertia, and any potential tidal heating \citep{selsis2013effect}. Many of these parameters are easier to constrain than the atmospheric circulation and cloud coverage of hot Jupiters, which means thermal phase curves of bare rocks are a promising way to test the tidal locking hypothesis.

Here we pursue this idea using observations of the super-Earth LHS 3844b. At a radius of 1.32 $R_{\bigoplus}$ and an orbital period of 11.1 hours, LHS 3844b is one of the smallest short period exoplanets with a measured thermal phase curve. The planet's Spitzer observations at 4.5 $\mu$m rule out many likely atmospheric scenarios and constrain the atmosphere's overall thickness to less than 1--10 bars \citep{kreidberg2019absence,whittaker2022}, compatible with the planet's transmission spectrum which rules out H$_2$-rich and H$_2$O-rich atmospheres in excess of 0.1 bars \citep{diamond-lowe2020a}. These observations are consistent with atmospheric escape models, which predict the planet lost its atmosphere due to the star's strong stellar wind and high XUV flux \citep{kreidberg2019absence,diamond-lowe2020,kane2020}. The most plausible interpretation is thus that LHS 3844b is a bare rock, which removes any potential degeneracy between non-synchronous rotation and atmospheric heat redistribution, and makes LHS 3844b ideal for testing the tidal locking hypothesis.

Previous analyses of LHS 3844b found that, besides ruling out a thick atmosphere, its phase curve is consistent with a circular orbit, synchronous rotation, and a relatively low-albedo surface composed of basalt \citep{kreidberg2019absence,whittaker2022}. Nevertheless, it is unclear whether such a state is necessarily the best fit to the data. A priori one can expect that LHS 3844b should be in a tidally evolved state, such as synchronous rotation, as the planet should no longer retain any initial rotation or eccentricity. We estimate it would only take $\sim 1.7\times 10^{10}$ sec $=$ 530 years for LHS 3844b's orbit to circularize\footnote{This estimate uses the tidal circularization timescale from \citet{puranam2018chaotic}, and a $Q$ value consistent with the tidal dissipation model in Equation \ref{eqn:tidal_dissipation}.}, with even less time required to spin down LHS 3844b's rotation. However, previous radial velocity measurements of LHS 3844 did not detect any periodic variations and only placed an upper bound of $\sim 1$ $M_{\rm Jup}$ on LHS 3844b's mass \citep{vanderspek2019tess}, which means the system could host one or more companion planets that could keep LHS 3844b's eccentricity excited and allow the planet to still be in non-synchronous rotation. Similarly, previous work showed that LHS 3844b's secondary eclipse requires the surface to be quite dark, with basalt a likely candidate for the surface's composition \citep{kreidberg2019absence,whittaker2022}. However, previous conclusions about LHS 3844b's surface composition were based on geologically fresh materials and did not address the potential impact of space weathering, even though space weathering is known to strongly modify the properties of atmosphere-less bodies in the Solar System \citep{pieters2016space}. The goal of this paper is thus to revisit the previous conclusions about LHS 3844b, and to explore what alternative scenarios might also fit the data.

The rest of this paper proceeds as follows. We develop a global thermal model for an atmosphere-less planet and compare its output to the Spitzer observations of LHS 3844b. The model is described in Section \ref{sec:method} and combines features of two previous exoplanet bare rock models \citep{hu2012theoretical,selsis2013effect}, including sub-surface heat diffusion, tidal heating, and bidirectional surface reflectance and emissivity. In contrast to previous studies of LHS 3844b, when comparing models against data we use the planet's full phase curve instead of just the derived secondary eclipse depth (see Section \ref{sec:method}). Section \ref{sec:results} finds that the Spitzer data rule out Mercury-like spin-orbit resonances as well as eccentricities larger than $e\gtrsim0.001$, largely because such spin and orbital states would cause intense tidal heating which is incompatible with the observations. In addition, we identify two scenarios which fit LHS 3844b's phase curve significantly better than a basalt planet on a circular orbit. The first requires that LHS 3844b still experiences weak tidal heating through a small-but-nonzero eccentricity (requiring a companion planet); the second requires that the surface of LHS 3844b has been substantially darkened by space weathering. Section \ref{sec:discussion} discusses why we cannot rule out tidal heating but consider space weathering more plausible, before Section \ref{sec:conclusion} concludes the paper.

\section{Methods}
\label{sec:method}

Our goal is to simulate the thermal phase curve of LHS 3844b for arbitrary rotational and orbital states. As long as the planet has no atmosphere, or the atmosphere is optically thin and does not significantly redistribute heat, the phase curve of a bare rock planet will be dominated by only a handful of processes, which we model as follows.

\subsection{Stellar flux}

For a given orbital configuration the incident stellar flux on the planet can be computed using Kepler's laws. Here we use the same approach as \citet{pierrehumbert2010principles}. The distance between star and planet is given by
\begin{eqnarray}
r (t) & = & a\frac{1-e^2}{1+e\cos (k_\theta(t))},
\end{eqnarray}
where $k_\theta(t)$ represents the planet-star orbital angle, $a$ the semi-major axis, and $e$ the planet's eccentricity. The change in $k_\theta(t)$ can be found from conservation of angular momentum, 
\begin{eqnarray}
\frac{\mathrm{d}k_\theta(t)}{\mathrm{d}t} & = & \frac{2\pi}{P}\frac{[1+e\cos(k_\theta(t))]^2}{(1-e^2)^{3/2}},
\end{eqnarray}
where $P$ denotes the planet's orbital period and $k_\theta(t)$ is numerically integrated to obtain the time-varying distance between the star and planet.

The flux absorbed by the planet at latitude $x$ and longitude $y$ is equal to
\begin{eqnarray}
    I_0(x,y,t)= \frac{R_s^2}{r(t)^2} \mu_0 \times \int (1-A(\lambda,\mu_0))I_s(\lambda)\mathrm{d}\lambda
    \label{absorption}
\end{eqnarray}
where $\lambda$ represents wavelength, $R_s$ the stellar radius, $I_s(\lambda)$ the stellar spectrum,  $A(\lambda,\mu_0)$ is a specific form of surface albedo called the directional-hemispherical reflectance (described in more detail below), and $\mu_0$ the zenith angle factor. The zenith angle factor is equal to $\mu_0=\cos(x-x_0(t))\cos(y-y_0(t))$ on the dayside and $\mu_0=0$ on the nightside, where $x_0$ and $y_0$ are the substellar latitude and longitude which are generally time-dependent. The above formulation assumes $R_s/r \ll 1$ so all incoming stellar flux is parallel \citep{nguyen2020}; LHS 3844b is roughly compatible with this assumption as $R_s/r \sim 0.14$. The above formulation also assumes LHS 3844b's obliquity is zero, justified by previous work which showed that obliquity decays quickly around low-mass stars \citep{barnes2008,heller2011}.

The planetary and stellar parameters are taken from \citet{vanderspek2019tess}, while the stellar spectrum of LHS 3844 is identical to that used in \citet{kreidberg2019absence}. The stellar spectrum was fitted in \citet{kreidberg2019absence} to the star's overall spectral energy distribution; however, its predicted flux of 55.9 mJy in the 4.5 $\mu m$ Spitzer bandpass did not match the observed flux of 43.4 mJy. Therefore a scaling constant was introduced to ensure that the stellar flux in the Spitzer bandpass matches the observed value. For consistency the same approach is used here --  we use the unscaled stellar spectrum to compute the surface temperature distribution of LHS 3844b, but multiply the stellar spectrum by a scaling constant of $55.9/43.4\sim1.3$ (applied to all wavelengths) when calculating the planet-star contrast ratio. Nonetheless, the best way of dealing with the mismatch between modeled and observed fluxes for LHS 3844 remains ambiguous, and one could also address it using alternative methods \citep{whittaker2022}.

\subsection{Tidal dissipation}
Planets with non-zero eccentricity or in spin-orbit resonances experience tidal dissipation which leads to internal heating. Here we model the resulting heat flux using the constant time lag model from \citet{leconte2010tidal}. The tidal dissipation rate is given by
\begin{eqnarray}
    E_{\rm tide}=2K_p[N_a(e)-2N(e)\frac{\omega}{n}+\Omega(e)\left(\frac{\omega}{n}\right)^2]
    \label{tide eq}
\end{eqnarray}
\noindent
where the tidal constants $N(e)$, $N_a(e)$, and $\Omega(e)$ describe the dependence of tidal dissipation on eccentricity, given by
\begin{eqnarray*}
        N(e)&=\frac{1+\frac{15}{2}e^2+\frac{45}{8}e^4+\frac{5}{16}e^6}{(1-e^2)^6},\\
        N_a(e)&=\frac{1+\frac{31}{2}e^2+\frac{255}{8}e^4+\frac{185}{16}e^6+\frac{25}{64}e^8}{(1-e^2)^{15/2}},\\
        \Omega(e)&=\frac{1+3e^2+\frac{3}{8}e^4}{(1-e^2)^{9/2}}.
\end{eqnarray*}
\noindent 
Here $n=2\pi/P$ represents the planet's mean motion, $\omega$ its angular velocity, and $K_p$ is a constant that represents the energy dissipated in each tidal cycle,
\begin{eqnarray}
    K_p=\frac32 k_p\Delta t_p \left(\frac{GM_p^2}{R_p}\right)\left(\frac{M_s}{M_p}\right)^2\left(\frac{R_p}{a}\right)^6n^2.
    \label{eqn:tidal_dissipation}
\end{eqnarray}
Above, $M_s$ and $M_p$ represent the star's and planet's mass, $R_s$ and $R_p$ represent the corresponding radii, $k_p$ is the Love number measuring the planet's rigidity and susceptibility to tidal potential \citep{love1909yielding}, and $\Delta t$ represents the time lag between maximum tidal potential and the planet's tidal bulge \citep{leconte2010tidal}.

The parameters $\Delta t$ and $k_p$ are typically fitted to observations, with $k_p=0.3$ and $\Delta t=630$ s providing a good match for present-day Earth's ocean-dominated tidal dissipation while $\Delta t \sim 60$ s is more appropriate for a rocky planet with tidal dissipation in its mantle only \citep{desurgy1997}. 
The internal structure of LHS 3844b is uncertain, including whether it might have an internal magma ocean. We therefore use Earth-like parameters as default for LHS 3844b, but we additionally explore uncertainty in tidal heating by increasing and decreasing $\Delta t$ by one order of magnitude.
If the tidal dissipation inside the planet is spatially uniform, then the surface heat flux is $e_{\rm tide}=E_{\rm tide}/4\pi R_p^2$.

For planets with non-zero eccentricity and outside of spin-orbit resonances, tidal dissipation is minimized in an orbital state called pseudo-synchronous rotation \citep{hut1981}. In this state the planet's rotation nearly matches its orbital period during apoastron, so the planet rotates faster over the course of an orbit than it would in synchronous rotation. The rotation rate for a pseudo-synchronous state is $\omega_{eq}/n = N(e)/\Omega(e)$, which simplifies the tidal dissipation equation to
\begin{eqnarray}
    e_{\rm tide}=2K_p[N_a(e)-\frac{N(e)^2}{\Omega(e)}]/(4\pi R_p^2).
\end{eqnarray}

Although it has been argued that rocky planets cannot be in pseudo-synchronous rotation because this state is incompatible with realistic rheological models of solids \citep{makarov2013}, we still consider pseudo-synchronous rotation for LHS 3844b because it provides a lower bound on the planet's tidal dissipation. If LHS 3844b was not in pseudo-synchronous rotation it would be captured into another steady state, such as synchronous rotation or a Mercury-like spin-orbit resonance (note, a planet in synchronous rotation with non-zero eccentricity will undergo Moon-like librations). For all of these states the planet's tidal dissipation would be even larger, which would only strengthen the constraints we derive from the planet's observed phase curve.

\subsection{Thermal diffusion}
Heat storage and heat redistribution inside the sub-surface are modeled using the thermal diffusion equation, $\partial T/\partial t = k/(c\rho) \nabla^2 T$,
% \begin{eqnarray}
%     \frac{\partial T}{\partial t} & = & \frac{k}{c\rho}\nabla^2T,
% \end{eqnarray}
% \noindent 
where the thermal conductivity $k$ and the volumetric heat capacity $c\rho$ can be combined into the thermal diffusivity $\alpha=k/c\rho$.
On planetary scales, horizontal heat transport due to heat diffusion can be ignored \citep{seager2009method}, so we only consider vertical heat diffusion,
\begin{eqnarray}
    \frac{\partial T}{\partial t}=\alpha \frac{\partial^2 T}{\partial z^2}.
\end{eqnarray}

To generate the planet's surface temperature map we numerically solve the thermal diffusion equation using the CLIMLAB climate modeling toolkit \citep{rose2018} on a $360\times 180 \times 10$ longitude-latitude-depth grid, where the depth points are linearly distributed between the surface and the model's bottom boundary. For planets not in synchronous rotation, we set the bottom boundary to 10 times the thermal skin depth \citep{spencer1989systematic}. For planets in synchronous rotation the thermal skin depth is infinite, so we set the bottom boundary to 10 times the thermal skin depth of a planet in 2:1 rotation. The top and bottom boundary conditions at each latitude-longitude point are
\begin{eqnarray}
    k\left(\frac{\partial T}{\partial z}\right)_{z=0}&=I_0-I_p, \\
    k\left(\frac{\partial T}{\partial z}\right)_{z=z_{\rm max}}&=-e_{\rm tide}.
\end{eqnarray}
Here $I_0$ is the absorbed stellar flux and $I_p$ is the emitted thermal flux
\begin{eqnarray}
    I_p=\int \varepsilon(\lambda)\pi B(\lambda,T_{z=0})\mathrm{d}\lambda.
    \label{emission}
\end{eqnarray}
where $B(\lambda,T)$ is the Planck function and $\varepsilon(\lambda)$ is the hemispherical emissivity (described below). Thermal observations of the Moon and Mercury can be fitted with a surface heat capacity $\Gamma=\sqrt{k\rho c}$ of about $50$ to $80$ J s$^{-1/2}$ K$^{-1}$ m$^{-2}$ \citep{1969ApJ...156.1135W,1974AJ.....79.1457M}, and other data suggest $\Gamma \sim 100$ J s$^{-1/2}$ K$^{-1}$ m$^{-2}$ for regolith surfaces \citep{pierrehumbert2010principles}, so here we use $\Gamma=100$ J s$^{-1/2}$ K$^{-1}$ m$^{-2}$. Tests show that our results are insensitive to the exact value of $\Gamma$ (see Appendix \ref{appendix:heatcapacity}).

\subsection{Bidirectional surface reflectance and emissivity}
Modeling the reflectance and emissivity of a bare rock surface poses a complex radiative transfer problem. Physically, an atmosphere-less body like the Moon or Mercury has a surface composed of loose unconsolidated material termed regolith (exoplanet surfaces composed of solid rock are possible but unlikely, given that space weathering via micrometeorites and stellar irradiation would convert these surfaces into regolith; see below). Stellar flux enters the regolith and then gets scattered, absorbed, and/or reemitted multiple times. This leads to a surface reflectance and emissivity that strongly depends on the surface's orientation with respect to the star and the observer, the regolith's composition, as well as interactions between regolith particles such as shadow-hiding and back-scattering \citep{hapke2002}.

Here we use a modeling approach originally developed by Hapke for Solar System bodies like the Moon and Mercury \citep{hapke1981a,hapke2002,hapke2012theory}, and which was subsequently applied to exoplanets \citep{hu2012theoretical}. In this approach one first infers the single-scattering albedo spectrum of a regolith particle based on laboratory reflectance measurements. The single-scattering albedo is then used in an analytical model which approximates the emergent flux of an illuminated semi-infinite regolith surface. Following \citet{hapke2012theory}, the surface's bidirectional reflectance function is given by
\begin{equation}
\begin{split}
    r(\lambda,\mu_0,\mu_e)=\frac{w(\lambda)}{4\pi}\frac{\mu_0}{\mu_0+\mu_e}[B_{S0}B_S(\mu_0,\mu_e)+\\H(\mu_0/K)H(\mu_e/K)].
    \label{eqn:bidirectional}
\end{split}
\end{equation}
Here $w(\lambda)$ is the wavelength-dependent single-scattering albedo, $\mu_0$ and $\mu_e$ are the cosine angles with which radiation enters and leaves the surface, $B_{S0}$ and $B_{S}$ are parameters that take into account the opposition effect, $H$ is an analytical function that arises from solving the radiative transfer equations, and $K$ primarily depends on the size of regolith particles; for comparison, the reflectance function for a Lambert surface is $r = w(\lambda)/(4\pi)$. We set $B_{S0}=0$ (no opposition effect) and $K=1$ to minimize the albedo at short wavelengths and maximize the observed secondary eclipse depth in the infrared. In addition, Equation \ref{eqn:bidirectional} assumes regolith particles scatter isotropically.

To solve the surface's energy balance one has to derive relevant albedo and emissivity quantities from the bidirectional reflectance function (Eqn.~\ref{eqn:bidirectional}). The relevant surface albedo in Equation~\ref{absorption} is the directional-hemispherical reflectance, 
\begin{equation}
%    A(\lambda,\mu_0)=\int_0^1 r(\mu_0,\mu_e,g)/\mu_0 \cdot 2\pi\mu_e d\mu_e
    A(\lambda,\mu_0)=\int_0^1 r(\lambda,\mu_0,\mu_e)/\mu_0 \cdot 2\pi\mu_e d\mu_e,
\end{equation}
which describes what fraction of a collimated light beam with incident angle $\mu_0$ is scattered back over all outgoing angles.
The relevant emissivity in Equation~\ref{emission} is the hemispherical emissivity $\varepsilon(\lambda)$, which describes the surface's emission over all outgoing angles. Applying Kirchhoff's law to the outgoing hemisphere, this quantity is
\begin{equation}
    \varepsilon(\lambda)=1-\int_0^1 A(\lambda,\mu_0)\cdot 2\pi\mu_0d\mu_0.
\end{equation}

We compute the above quantities using single-scattering albedos that were derived based on laboratory measurements in \citet{hu2012theoretical}.
To validate our model we compared it to a set of bidirectional reflectance simulations performed with the numerical model from \citet{hu2012theoretical}. The two models agree fairly closely, with some remaining model differences that could be either due to round-off errors in fundamental constants or different choices in model resolution (see Appendix \ref{validation}).

%To validate our model we compare its bidirectional reflectance simulations to the simulations presented in \citet{hu2012theoretical}. Our model closely reproduces the previous calculations in the parameter space relevant for LHS 3844b, even though we find some remaining differences for colder planets which could be due to differences in the assumed stellar model (see Appendix \ref{validation}).

\subsection{Space weathering}
Space weathering refers to alterations that occur in the space environment over time \citep{pieters2016space}.  On an exposed surface, impacts by micrometeorites and stellar irradiation lead to comminution and melting. Besides reworking the surface and generating a regolith, these alterations can also cover the surface with molten ferric material called nanophase metallic iron. Similarly, micrometeorites can deliver exogeneous material and thus modify the composition of the surface's top layer \citep{syal2015darkening}.
In the Solar System these processes tend to lower the albedo of atmosphere-less bodies and modify their spectral signatures, so they should also be considered for atmosphere-less exoplanets. The Moon's low surface albedo is due to weathering by nanophase metallic iron \citep{cassidy1975,pieters2000}, while Mercury's low surface albedo has been attributed to mixed-in graphite \citep{syal2015darkening}, so we consider both possibilities here.

We first assume LHS 3844b's surface is primarily composed of basalt. Previous analyses of LHS 3844b showed that its secondary eclipse depth can be variously matched by basalt, pyrite (also previously called metal-rich), or a mixture of basalt and nanophase hematite (also called Fe-oxidized) \citep{kreidberg2019absence,whittaker2022}. Of these, basalt is the most plausible candidate \citep{hu2012theoretical}; the lunar mare are basaltic \citep{pieters1978}, while the other surface types proposed for LHS 3844b are not found on atmosphere-less bodies in the Solar System and so would require invoking unusual formation scenarios.

Next, to simulate how space weathering modifies the surface's spectral properties we follow the same approach as \citet{hapke2001space} and \citet{hapke2012theory}. For simplicity we only consider uniform surfaces, so space weathering affects all latitude-longitude points equally. At a given point, the surface is assumed to consist of a mixture of host material (basalt) and absorbing particles. One then computes an effective single-scattering albedo for the mixture as follows. The relationship between the absorption coefficient $\alpha$ and the single scattering albedo $w$ of either host material or mixture is 
\citep{hapke2012theory}
\begin{eqnarray}
    \alpha_h=\frac{1}{D}\ln[S_t+\frac{(1-S_e)(1-S_t)}{w_h-S_e}]\\
    \alpha_w=\frac{1}{D}\ln[S_t+\frac{(1-S_e)(1-S_t)}{w_w-S_e}]
    \label{Hapke}
\end{eqnarray}
where subscript $h$ refers to the host material, subscript $w$ refers to the mixture, $D$ is the mean photon path length through the host material, $w_h$ and $w_w$ are the single scattering albedos of the host material and mixture, and $S_e$ and $S_t$ are angle-averaged Fresnel reflection coefficients of the host material \citep[see][]{hapke2001space}.
The single scattering albedo of the mixture $w_w$ can then be solved for using 
\begin{eqnarray}
    \alpha_w=\alpha_h+36\pi z\phi/\lambda
\end{eqnarray}
\noindent
where $\phi$ is the volume fraction of absorbing particles, and $z$ is related to the refractive indices of host and absorbing material, defined as
\begin{eqnarray}
    z=\frac{n_h^3 n_{m} k_{m}}{(n_{m}^2-k_{m}^2+2n_h^2)^2+(2n_{m} k_{m})^2}
\end{eqnarray}
where $n_h$ is the real part of the host material's refractive index, while $n_{m}$ and $k_{m}$ are the real and imaginary parts of the absorbing material's refractive index \citep{hapke2001space,hapke2012theory}. The refractive indeces of graphite and iron nanoparticles are evaluated using data from the Refractive Index database \citep{rii2022}.

\subsection{Comparison against observations}

In comparing models against data for LHS 3844b there are two options. Previous work compared various physical models to the best-fit secondary eclipse depth of $380 \pm 40$ ppm (all error bars in this work are 1$\sigma$ uncertainty), which itself was derived from the observed Spitzer phase curve \citep{kreidberg2019absence,whittaker2022}. However, we find that the previously-derived secondary eclipse depth only provides a modest constraint for discriminating between models, as all surface models we consider in detail match this depth to within 2$\sigma$ (see below).
Moreover, the best-fit secondary eclipse depth of $380 \pm 40$ ppm reported in \citet{kreidberg2019absence} was based on a phase curve model with a day-night flux amplitude of $350 \pm 40$ ppm, and thus implicitly requires a best-fit nightside flux slightly larger than zero (although the data also matches zero nightside flux within $1 \sigma$).
A second option is to directly compare models to the observed Spitzer phase curve based on $\chi^2$. Doing so allows us to include all out-of-eclipse data points in the model-data comparison, and moreover avoids any assumptions about the planet's nightside flux which are implicit when using the derived secondary eclipse depth from \citet{kreidberg2019absence}. When reporting $\chi^2$ and $\chi^2_\nu$ we use the binned phase curve from \citet{kreidberg2019absence}, which covers 24 orbital phase bins and is normalized with respect to the secondary eclipse, for $\nu=23$ degrees of freedom.

To simulate phase curves we assume an edge-on orbit. For non-circular orbits one additionally needs to specify when the secondary eclipse occurs relative to periastron. We assume LHS 3844b's periastron coincides with secondary eclipse (orbital phase of $0.5$) and its apoastron coincides with transit (orbital phase of $0$). We do not consider alternative viewing geometries as our thermal modeling analysis constrains the planet's eccentricity to less than $\sim 10^{-3}$ (see below), for which the impact of viewing geometry is negligible. Neither do we attempt to constrain LHS 3844b's eccentricity from its observed transit duration or the timing between secondary eclipse and transit, as eccentricity constraints from these methods are typically much less precise \citep[][]{vaneylen2015,sagear2023} than the constraint we derive from thermal modeling.

The planet's phase curve is composed of reflected and emitted light. Unlike the angle-integrated albedo and emissivity quantities that are used to compute the surface's energy balance (see above), the observed albedo and emissivity of a given latitude-longitude patch also depend on the patch's orientation with respect to the observer. We therefore compute observed reflected and emitted light components using the bidirectional reflectance function (Eqn.~\ref{eqn:bidirectional}), which is separately applied to each visible latitude-longitude patch on the planet's surface.

Finally, to compute the observed planet-star flux contrast in the Spitzer bandpass we use a spectral average weighted by the Spitzer response function \citep{instrument2013instrument}.
Note that in computing photometric contrasts the order of the spectral averaging procedure matters. Here we spectrally average fluxes first before taking their ratio
\begin{eqnarray}
    \mathrm{contrast} & = & \frac{\overline{F_p(\lambda)}}{\overline{F_s(\lambda)}} = \frac{\int F_p(\lambda) f(\lambda) d\lambda}{\int F_s(\lambda) f(\lambda) d\lambda},
\end{eqnarray}
which is not the same as the alternative of taking the ratio first before spectrally averaging, $\overline{F_p(\lambda)/F_s(\lambda)} \neq \overline{F_p(\lambda)}/\overline{F_s(\lambda)}$. Above overbars denote a weighted spectral average over the Spitzer bandpass, $F_p$ and $F_s$ are the observed planet and star fluxes, and $f(\lambda)$ is the Spitzer response function.

%which is not the same as the alternative but incorrect procedure of taking the ratio first before spectrally averaging, $\overline{F_p(\lambda)/F_s(\lambda)} \neq \overline{F_p(\lambda)}/\overline{F_s(\lambda)}$. Above overbars denote a weighted spectral average over the Spitzer bandpass, $F_p$ and $F_s$ are the observed planet and star fluxes, and $f(\lambda)$ is the Spitzer response function.

\begin{figure*}[ht]
    \centering
    \includegraphics[width=0.7\textwidth]{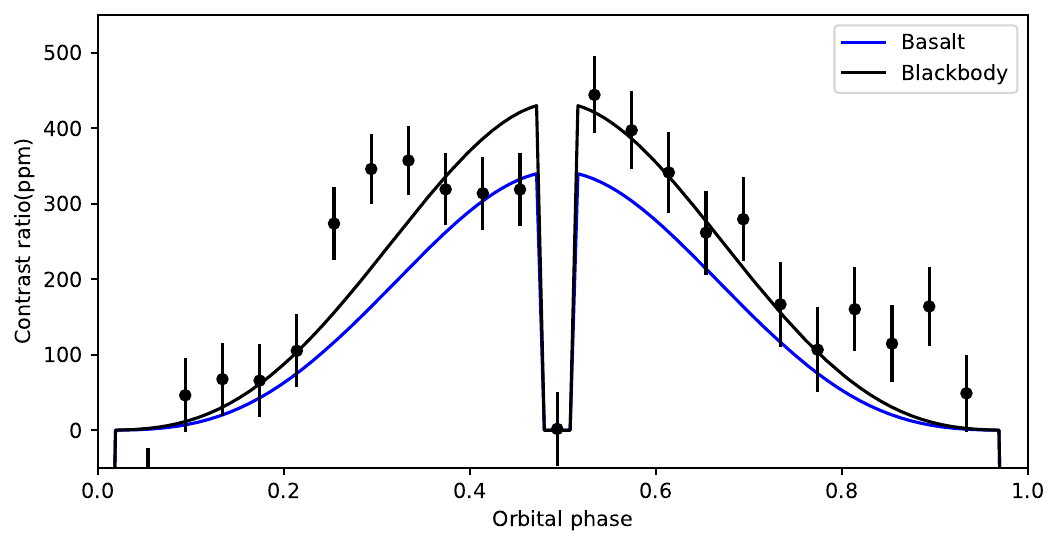}
    \caption{Phase curve of LHS3844b. Dots show Spitzer data, lines are bare rock models which assume the planet is synchronously rotating and on a circular orbit.}
    \label{fig:bare rock}
\end{figure*}

\section{Results}
\label{sec:results}

\subsection{Previous conclusions}
\label{sec:null_hypothesis}
We first revisit the conclusion of \citet{kreidberg2019absence} that the secondary eclipse of LHS 3844b is explained well by a planet in synchronous rotation, with zero eccentricity, and a basalt surface.
Figure \ref{fig:bare rock} shows that although this scenario is consistent with the planet's secondary eclipse, it does not provide the best fit to the overall phase curve. Focusing on the secondary eclipse, a pure basalt surface would have an eclipse depth of $339$ ppm, which compares well against the previously-derived secondary eclipse depth of $380 \pm 40$ ppm (see Table \ref{tab:chi2 compare}). For comparison, a hypothetical blackbody would produce a secondary eclipse depth of $434$ ppm, which also matches the previously-derived secondary eclipse depth to better than 2$\sigma$. Nevertheless, the fit to the overall phase curve is notably different between these two models. For the basalt surface we find a fit of only $\chi^2 = 75.78$ ($\chi^2_\nu = 3.3$, p-value = $1.5\times 10^{-7}$), whereas a blackbody would have a much better fit of $\chi^2 = 44.0$ ($\chi^2_\nu = 1.9$, p-value = $5\times 10^{-3}$). The difference is visible in Figure \ref{fig:bare rock}, as a basalt surface matches the observed fluxes clearly worse than a blackbody, particularly around quadrature (near orbital phases $0.25$ and $0.75$).
%While a synchronously rotating planet with a basalt surface is thus able to match LHS 3844b's reported secondary eclipse depth, its poor fit to the planet's phase curve prompts us to explore alternative hypotheses that could also match the data.

Based on Figure \ref{fig:bare rock}, one might further expect that an asymmetric surface model could fit the data even better than a symmetric one; the observed flux immediately before secondary eclipse is $\sim$50-100 ppm lower than the flux right after eclipse. However, the apparent dayside asymmetry is likely an artifact. First, previous work found that the data bin down as expected based on photon noise statistics, whereas a robust asymmetry should have led to a noticeable deviation from photon noise at large bin sizes \citep[see Extended Data Fig 3 in][]{kreidberg2019absence}. Second, \citet{kreidberg2019absence} reported that the best phase curve fit was consistent with no hot spot offset, again supporting symmetric models. In this work we therefore focus on models with uniform surface properties, such that any phase curve asymmetry only enters through the surface's heat capacity (in practice this effect is negligible; see Fig.~\ref{tide effect}).

Overall, while a synchronously rotating planet with a basalt surface is thus able to match LHS 3844b's reported secondary eclipse depth, its poor fit to the planet's phase curve prompts us to explore alternative hypotheses that could also match the data.

%%%---
%We first revisit the conclusion of \citet{kreidberg2019absence} that the secondary eclipse of LHS 3844b is explained well by a planet in synchronous rotation, with zero eccentricity, and a basalt surface.

%Figure \ref{fig:bare rock} shows that although this scenario is consistent with the planet's secondary eclipse, it does not provide the best fit to the overall phase curve. Focusing on the secondary eclipse, a pure basalt surface would have an eclipse depth of $339$ ppm, which compares well against the previously-derived secondary eclipse depth of $380 \pm 40$ ppm (see Table \ref{tab:chi2 compare}). For comparison, a hypothetical blackbody would produce a secondary eclipse depth of $434$ ppm, which also matches the previously-derived secondary eclipse depth to better than 2$\sigma$. Nevertheless, the fit to the overall phase curve is notably different between these two models. For the basalt surface we find a fit of only $\chi^2 = 75.78$ ($\chi^2_\nu = 3.3$, p-value = $1.5\times 10^{-7}$), whereas a blackbody would have a much better fit of $\chi^2 = 44.0$ ($\chi^2_\nu = 1.9$, p-value = $5\times 10^{-3}$). The difference is visible in Figure \ref{fig:bare rock}, as a basalt surface matches the observed fluxes clearly worse than a blackbody, particularly around quadrature (near orbital phases $0.25$ and $0.75$). Therefore, while a synchronously rotating planet with a basalt surface is able to match LHS 3844b's reported secondary eclipse depth well, its poor fit to the planet's phase curve prompts us to explore alternative hypotheses that could also match the data.

\subsection{Non-zero eccentricity and non-synchronous rotation}

\begin{figure*}[ht]
    \centering
    \includegraphics[width=0.6\textwidth]{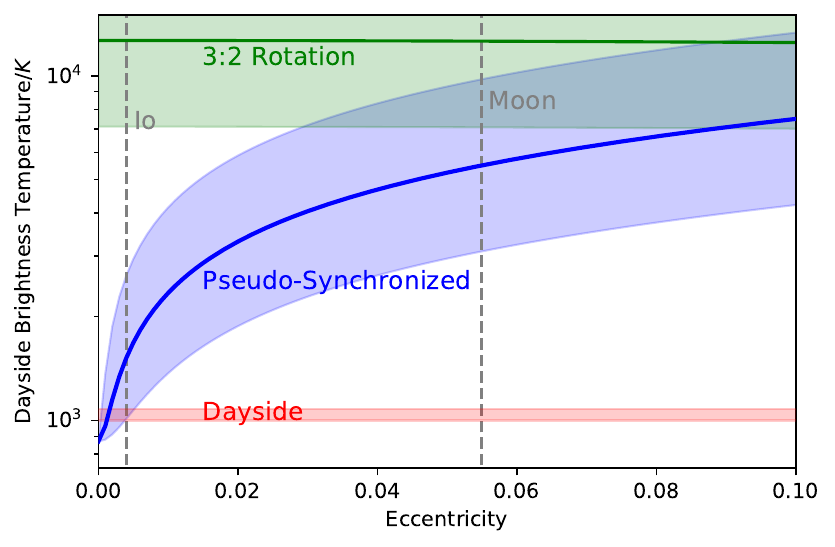}
    \caption{Observed dayside brightness temperature of LHS3844b (red; $1 \sigma$ range), versus the planet's theoretical dayside equilibrium temperature for two different rotation states (solid curves). 
    The blue curve assumes pseudo-synchronous rotation, the green curve assumes Mercury-like 3:2 rotation. Shaded blue and green regions indicate the effect of increasing or decreasing LHS 3844b's tidal dissipation efficiency by one order of magnitude.
    %All the cases assume the planet is lambertian scattering. 
    }
    \label{fig:brightness}
\end{figure*}

Next, we revisit LHS 3844b's rotation and eccentricity. To establish an upper bound for both parameters, we compare the observed dayside brightness temperature of $1040\pm40$ K \citep{kreidberg2019absence} with the planet's theoretical dayside equilibrium temperature $T_{\rm day}$. Tidal dissipation resulting from non-zero eccentricity or non-synchronous rotation will cause the equilibrium temperature to rise and possibly exceed the observed dayside brightness temperature, which we use here to limit the plausible parameter space.

The dayside equilibrium temperature, incorporating tidal dissipation, is equal to %$\sigma T_{day}^4=e_{\rm tide} + (R_s/(\sqrt{2}\overline{r}))^2 \int (1-\alpha) I_s d\lambda$.
%$\sigma T_{day}^4=e_{\rm tide} + (R_s/(\sqrt{2}\overline{r}))^2 \int \int (1-A) \mu_0 I_s d\lambda d\mu_0$.
\begin{eqnarray}
    \sigma T_{\rm day}^4 & =& e_{\rm tide} + \frac{R_s^2}{2 \overline{r}^2} \int \int (1-A) \mu_0 I_s d\lambda d\mu_0.
\end{eqnarray}
This form assumes a uniform dayside and no heat redistribution to the nightside. The surface albedo is computed using the single-scattering albedo of basalt (for simplicity here we replace $A$'s angular dependence with Lambert scattering), $\overline{r}$ is the time-averaged planet-star distance, and $e_{\rm tide}$ is the tidal dissipation, which is a function of planetary eccentricity and rotation rate. 
As $e_{\rm tide}$ also depends on the planet's internal structure, which is uncertain, we vary the tidal time lag parameter $\Delta t$ (see Eqn.~\ref{eqn:tidal_dissipation}) by one order of magnitude upwards and downwards.

Figure \ref{fig:brightness} shows the planet's dayside equilibrium temperature $T_{\rm day}$ for a planet that is either pseudo-synchronized (blue curve) or in a Mercury-like 3:2 spin-orbit resonance (green curve). The colored regions indicate the impact of varying $\Delta t$ by one order of magnitude. We find that, regardless of the exact value of $\Delta t$, if LHS 3844b were in a 3:2 resonance, tidal heating would raise the dayside temperature to over 7,000 K which is significantly higher than its host star temperature. This large level of heating is inconsistent with both the Spitzer data and the lack of stellar-temperature companions in the original TESS phase curve \citep{vanderspek2019tess}.

Additionally, Figure \ref{fig:brightness} constrains the planet's eccentricity. Assuming LHS 3844b is in pseudo-synchronous rotation and for the default value of $\Delta t$ (solid blue curve), the eccentricity must be smaller than 0.002 %0.0021
for $T_{\rm day}$ to be within $3\sigma$ of the dayside brightness temperature. Even if $\Delta t$ is decreased by one order of magnitude to allow for slightly larger eccentricities, only eccentricities below 0.005 %0.0053
are permitted. For comparison, the Moon and Io have eccentricities of 0.05 and 0.004, respectively \citep{meeus1997mathematical,Peale1979}.
This estimate is conservative as it assumes pseudo-synchronous rotation. As explained above, if LHS 3844b was in synchronous rotation the resulting tidal dissipation would be higher, and the permitted eccentricity lower. Even without any further analysis, the absence of large observed tidal dissipation on LHS 3844b thus strongly rules out Mercury-like rotation states and constrains its orbit to be about as circular as Io's.

\begin{figure*}[h]
    \centering 
    \includegraphics[width=0.7\textwidth]{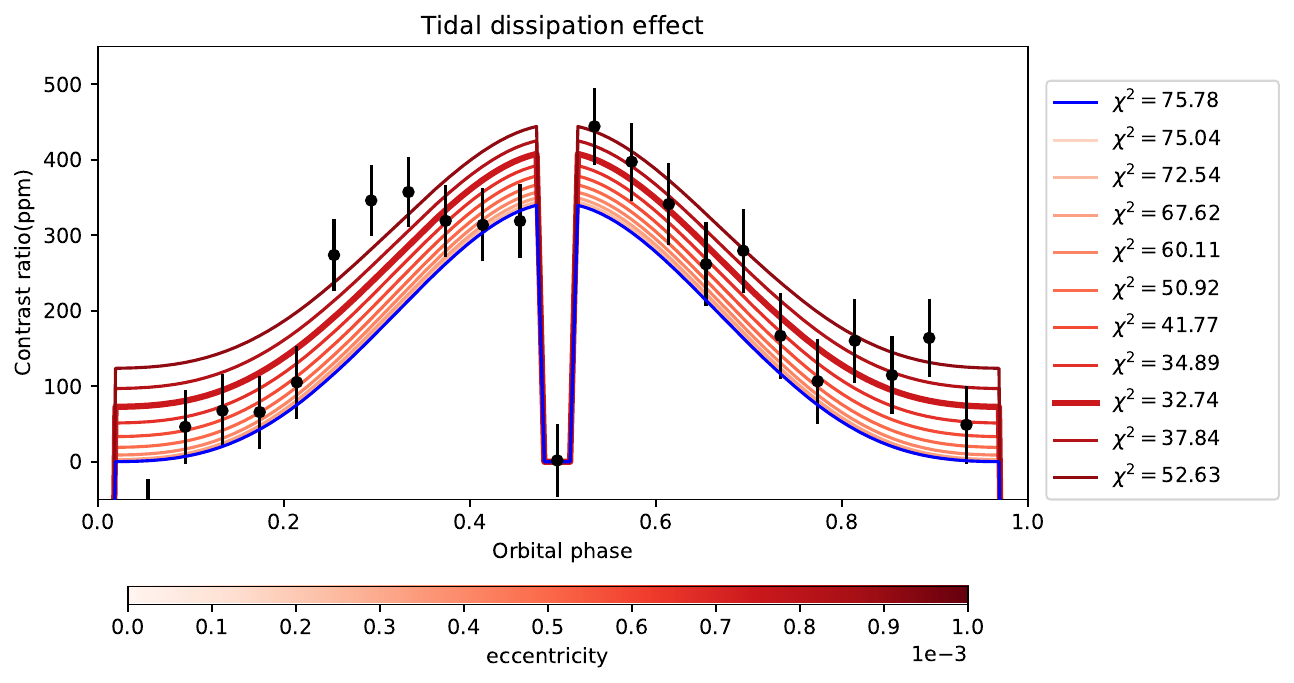}
    \caption{Simulated phase curves as a function of eccentricity, assuming LHS 3844b is in pseudo-synchronous rotation and has a basalt surface. Blue curve shows zero eccentricity and is identical to the curve in Fig.~\ref{fig:bare rock}.
    } 
    \label{tide effect}
\end{figure*}

For the non-zero eccentricities consistent with LHS 3844b's dayside brightness temperature in Figure \ref{fig:brightness}, which one provides the best fit to the Spitzer phase curve?
To address this question we use our thermal model to simulate detailed latitude-longitude temperature maps and phase curves. The calculations assume the planet is pseudo-synchronized, its surface composition is basalt, and $\Delta t= 630$ s. Figure \ref{tide effect} shows that as eccentricity increases, the flux curves all rise in response to increasing tidal dissipation.
If we broadly consider phase curves with $\chi_\nu^2 < 2$ acceptable, so their fit is at least as good as that of a zero-eccentricity blackbody (see Table~\ref{tab:chi2 compare}), then the acceptable eccentricity range for LHS3844b is $4\times10^{-4} < e < 10^{-3}$, or less than 25\% of Io's eccentricity. Within this eccentricity range LHS 3844b’s rotation has to be slower than $\omega/n < $1.000006, or about one planet rotation every 200 years. This means even if LHS 3844b was in pseudo-synchronous rotation, its rotation would be so slow as to be effectively tidally locked.

Figure \ref{tide effect} additionally shows that, for a surface made of basalt, LHS 3844b's phase curve also requires tidal dissipation. The best fit occurs at $e=0.0008$, that is, at non-zero eccentricity and tidal heating. The tidal heat flux associated with this eccentricity is about 10700 W m$^{-2}$, significantly smaller than LHS 3844b's stellar constant of about 95600 W m$^{-2}$. The resulting fit, $\chi^2=32.74$ ($\chi^2_\nu=1.4$, p-value=$0.09$), is significantly better than the above-described fit for a planet with a basalt surface and zero eccentricity, $\chi^2=75.78$ (see Table~\ref{tab:chi2 compare}). Based on LHS 3844b's extremely short tidal circularization timescale of $\sim 500$ years (see Section \ref{sec:introduction}), a non-zero eccentricity for LHS 3844b almost certainly requires the presence of a planetary companion. We discuss this possibility in Section \ref{sec:discussion}.
However, a surface made of pure basalt might not provide the best fit to LHS 3844b's phase curve either, so we revisit the planet's surface composition next.

\subsection{Space Weathering}

\begin{figure*}[th]
    \centering
    \includegraphics[width=\textwidth]{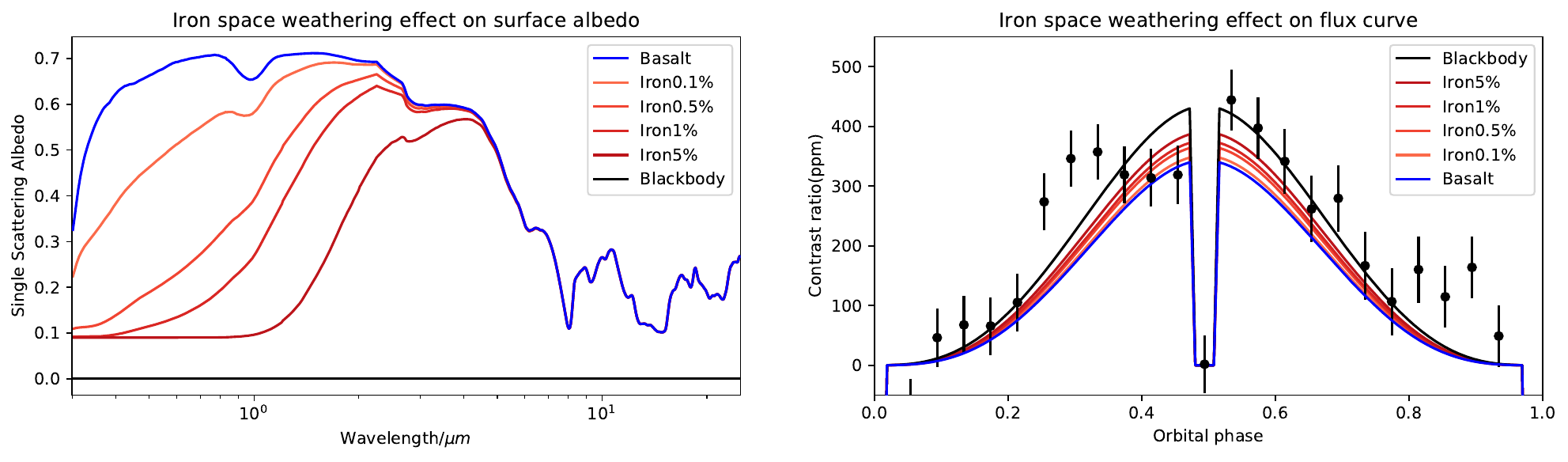}\\
    \includegraphics[width=\textwidth]{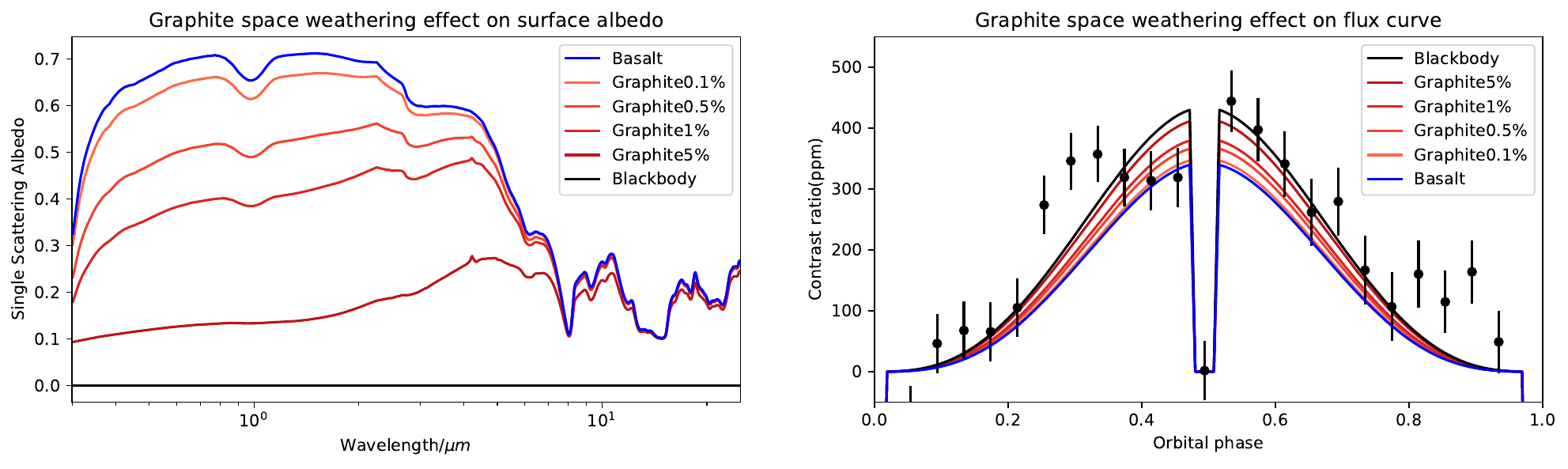}
    \caption{Effect of space weathering on the single scattering albedo of basalt (left column), and resulting thermal phase curves (right column). Top row shows the effect of space weathering via the formation of nanophase metallic iron particles (npFe), bottom row shows the effect of space weathering via graphite. Percentages show the added volume fraction of darkening material.
    }
    \label{fig:space_weathering}
\end{figure*}

Motivated by the Moon and Mercury, we consider how space weathering via nanophase metallic iron particles (npFe) and graphite would affect LHS 3844b's surface. What is the maximum impact these processes could have? If space weathering primarily reduces surface iron into npFe \citep{domingue2014}, then an upper bound for LHS 3844b is the limit in which all surface iron has been reduced. Earth's crust contains $\sim 5\%$ iron, which is on the same order of magnitude as the surface iron abundances of the Moon, Mercury, and Mars \citep{fleischer1954,lucey1995,robinson2001,taylor2006}. A highly space-weathered surface with a composition similar to Solar System bodies can therefore reasonably contain up to $\sim 5\%$ npFe. Correspondingly, we compute the surface albedo of LHS 3844b for mixtures ranging from an unweathered surface (pure basalt) up to highly weathered surfaces with $5\%$ volume fraction of either npFe or graphite.

The left column of Figure \ref{fig:space_weathering} shows the single scattering albedo spectrum for different amounts of space weathering. Regardless of whether space weathering occurs via npFe particles or graphite, albedo is always dramatically reduced at wavelengths shorter than 5 $\mu$m. The change depends non-linearly on the amount of darkening material -- for example, at 1 $\mu$m only $0.5\%$ of added npFe is necessary to reduce the surface's single scattering albedo by half, but the effect saturates above $5\%$ npFe. The non-linearity arises because small amounts of a dark absorber inside a reflective medium have a nonlinear effect on the mixture's overall absorption \citep{hapke2001space}. Per volume fraction of absorbing material, graphite is more effective at reducing albedo than npFe particles. In the near-infrared the surface's single scattering albedo still exceeds $0.5$ with $5\%$ of npFe added, whereas the single scattering albedo is lowered to less than $0.3$ at all wavelengths with $5\%$ of graphite.

The right column of Figure \ref{fig:space_weathering} shows the corresponding thermal phase curves, assuming LHS 3844b has zero eccentricity and is synchronously rotating. Because space weathering reduces the surface albedo, it always increases the phase curve amplitude. In all cases the largest flux originates at the substellar point, consistent with fits to the Spitzer phase curve which constrain the hot spot to lie within $-6\pm 6$ degrees of the substellar point \citep{kreidberg2019absence}.

Compared to a pure basalt surface, we find that models with space weathering are essentially indistinguishable in terms of their fit to the previously-derived secondary eclipse depth, but they significantly improve the fit to the overall phase curve. Table \ref{tab:chi2 compare} shows that all models with space weathering match the secondary eclipse depth of $380 \pm 40$ ppm to better than $2 \sigma$.
In contrast, the fit to the phase curve improves monotonically with the addition of space weathering. For reference, for a pure basalt surface $\chi^2=75.78$ (see Table~\ref{tab:chi2 compare}). Strong space weathering via the addition of $5\%$ npFe improves the phase curve fit to $\chi^2=53.07$ ($\chi^2_\nu=2.4$, p-value = $4 \times 10^{-4}$), while strong space weathering with $5\%$ graphite produces an even better fit, $\chi^2=46.48$ ($\chi^2_\nu=2.0$, p-value = $3 \times 10^{-3}$). Note that the latter model with $5\%$ graphite produces a phase curve that is almost identical to that of a blackbody ($\chi^2=46.48$ versus $\chi^2=44.0$; see Figure \ref{fig:space_weathering}). We conclude that, similar to tidal heating, the addition of space weathering significantly improves the fit to LHS 3844b's phase curve.

\begin{table*}[th!]
    \centering
    \caption{Secondary eclipse depth and $\chi^2$ phase curve fit for different surface compositions. All models match the previously-derived secondary eclipse depth of $380 \pm 40$ ppm to within 2$\sigma$ \citep{kreidberg2019absence}, but differ in their fits to the overall phase curve. Bold marks models that minimize $\chi^2$.
    Center columns use the original error bars from \citet{kreidberg2019absence}, while right columns inflate error bars by 1.383 to achieve $\chi^2_\nu=1$ for a black body.
    P-values show the probability of obtaining a $\chi^2$ value at least as large as the one observed, assuming the respective model is correct.
    }
    \label{tab:chi2 compare}
    \begin{tabular}{lccccc}
    \hline
      \multirow{2}*{Surface} & Secondary eclipse&\multicolumn{2}{c}{Phase curve, original error}&\multicolumn{2}{c}{Phase curve, inflated error}\\
      &(ppm)&$\chi^2$&p-value&$\chi^2$&p-value\\
      \hline\hline
      Derived from phase curve \citep{kreidberg2019absence} & $380\pm 40$& & & \\
      Black body & 434 & \textbf{44.00} & \textbf{$5.26\times 10^{-3}$} & \textbf{23.00} & \textbf{$0.460$}\\
      Basalt  & 343 & 75.78 & $1.51\times 10^{-7}$ & 39.62 & $0.017$\\
      Basalt, with tidal dissipation ($e$=0.0002) &347&72.54&$4.90\times 10^{-7}$&37.93&0.026\\
      Basalt, with tidal dissipation ($e$=0.0004) &360&60.11&$3.68\times 10^{-5}$&31.43&0.113\\
      Basalt, with tidal dissipation ($e$=0.0006)
      &382&41.77&$9.66\times 10^{-3}$&31.84&0.530\\
      Basalt, with tidal dissipation ($e$=0.0008) &411&\textbf{32.74}&$8.57\times 10^{-2}$&\textbf{17.12}&0.803\\
      Basalt, with tidal dissipation ($e$=0.0010) &447&52.63&$4.11\times 10^{-4}$&27.52&0.235\\
      Basalt, with iron space weathering 0.1\% & 351 & 71.10 & $8.20\times 10^{-7}$ & 37.17 & $0.031$\\
      %Iron space weathering 0.2\% & 357 & 68.07 & $2.41\times 10^{-6}$ & 35.59 & $0.045$\\   
      %Iron space weathering 0.3\% & 361 & 65.89 & $5.14\times 10^{-6}$ & 34.45 & $0.058$\\
      Basalt, with iron space weathering 0.5\% & 367 & 62.93 & $1.42\times 10^{-5}$ & 32.90 & $0.083$\\
      Basalt, with iron space weathering 1\% & 376 & 59.01 & $5.30\times 10^{-5}$ & 30.85 & $0.126$\\
      %Iron space weathering 2\% & \textbf{383} & 55.81 & $1.51\times 10^{-4}$ & 29.18 & $0.174$\\
      %Iron space weathering 3\% & 387 & 54.39 & $2.36\times 10^{-4}$ & 28.44 & $0.200$\\
      Basalt, with iron space weathering 5\% & 390 & \textbf{53.07} & \textbf{$3.58\times 10^{-4}$} & \textbf{27.75} & \textbf{$0.225$}\\
    %   Iron space weathering 10\% & 417 & 1.37\\
    %   Iron space weathering 20\% & 423 & 1.35\\
      Basalt, with graphite space weathering 0.1\% & 350 & 71.50 & $7.12\times 10^{-7}$ & 37.38 & $0.030$\\
      %Graphite space weathering 0.2\% & 356 & 68.18 & $2.32\times 10^{-6}$ & 35.64 & $0.045$\\
      %Graphite space weathering 0.3\% & 361 & 65.55 & $5.80\times 10^{-6}$ & 34.27 & $0.061$\\
      Basalt, with graphite space weathering 0.5\% & 370 & 61.64 & $2.21\times 10^{-5}$ & 32.23 & $0.096$\\
      Basalt, with graphite space weathering 1\% & 383 & 55.91 & $1.46\times 10^{-4}$ & 29.23 & $0.173$\\
      %Graphite space weathering 2\% & 398 & 50.81 & $7.20\times 10^{-4}$ & 26.57 & $0.275$\\
      %Graphite space weathering 3\% & 406 & 48.51 & $1.44\times 10^{-3}$ & 25.36 & $0.332$\\
      Basalt, with graphite space weathering 5\% & 415 & \textbf{46.48} & \textbf{$2.61\times 10^{-3}$} & \textbf{24.30} & \textbf{$0.387$}\\
    %   Graphite space weathering 10\% & 425 & 1.35\\
    %   Graphite space weathering 20\% & 428 & 1.35\\
     \hline
    \end{tabular}
\end{table*}

\section{Discussion}
\label{sec:discussion}

Our results show a pure basalt surface with zero tidal heating provides a poor fit to LHS 3844b's phase curve ($\chi^2_\nu = 3.3$), while the fit is significantly improved if we include either small tidal heating associated with a residual eccentricity (best model, $\chi^2_\nu = 1.4$), or space weathering (best model, $\chi^2_\nu = 2.0$). Of these two hypotheses, which one is more likely?

If we take the observational errors from \citet{kreidberg2019absence} at face value then tidal heating provides a better fit to the data. One way to evaluate the different models in Table \ref{tab:chi2 compare} is to impose a p-value cutoff. Considering only models with p-value above $0.05$ as acceptable ($\chi^2 < 35.17$), the only model that provides a reasonable fit to the data is the model with tidal heating and $e=0.0008$.
As an alternative, one can also use the minimum $\chi^2$ method \citep{avni1976}. Treating the degree of tidal dissipation and space weathering in our thermal model as two independent parameters, one can reject with $0.99$ significance level (roughly $3\sigma$) all models that have $\Delta \chi^2 > 9.21$ relative to the model which minimizes $\chi^2$. Again, we find that in this case only models with tidal heating fall inside the $0.99$ significance level relative to the best-fitting model with $e=0.0008$.
Based on LHS 3844b's short tidal circularization timescale, any non-zero tidal heating for the planet would require a third body that keeps its orbit excited, such as a planetary companion. Given the large uncertainties in the system's radial velocity data, a companion to LHS 3844b cannot be ruled out \citep{vanderspek2019tess}. More precise radial velocity measurements of LHS 3844 would thus be highly desirable.

Nevertheless, we do not consider tidal heating necessary to explain LHS 3844b's phase curve. Tidal heating requires one to invoke a still-undiscovered planetary companion, whereas by analogy with the Solar System space weathering is highly likely to have occurred on LHS 3844b.
As a pessimistic estimate we inflate the observational errors from \citet{kreidberg2019absence} by 38.3$\%$, so a blackbody would result in a phase curve fit of $\chi^2_\nu=1$ (see Table \ref{tab:chi2 compare}, right columns).
In this case a wide range of models with tidal heating and space weathering are able to match the data well, both based on their absolute p-values and $\Delta \chi^2$. Note that even with inflated error bars a pure basalt surface without tidal heating provides a poor fit to the data (p-value=$0.02$, $\Delta \chi^2 = 22.5$, so $>3\sigma$ worse compared to the best-fitting model).
We infer that LHS 3844b's phase curve indicates either tidal heating or space weathering. The first of these hypotheses requires strong assumptions about the existence of an additional planet, whereas the second hypothesis is physically plausible and can also match the data well if previous studies slightly underestimated the true observational error \citep[also see][]{hansen2014}.

Finally, to what extent are our conclusions affected by modeling uncertainty? 
The main conclusion, namely that LHS 3844b has to be in synchronous rotation (or nearly so) and has to have a small eccentricity, is sensitive to the assumed tidal model. In the constant time lag model uncertainty about the planet's internal structure is parameterized by the parameter $\Delta t$ \citep{leconte2010tidal}, but alternative modeling approaches exist. We tested the tidal parameterization from the VPLanet model \citep{barnes2020}, and found tidal heat fluxes consistent with the constant time lag model to within a factor of two. Nevertheless, the planet's tidal heating could vary more strongly than anticipated. For example, tidal heating on Io is thought to generate a shallow subsurface magma layer \citep{khurana2011,tyler2015}. Depending on internal heat sources, LHS 3844b might similarly sustain an internal magma ocean which could strongly affect the planet's tidal response. Future work with more detailed tidal models is thus desirable to support our conclusions.

\section{Conclusion}
\label{sec:conclusion}

We have re-analyzed the Spitzer thermal phase curve of LHS 3844b using a global bare rock thermal model. We find that the Spitzer data rule out non-synchronous rotation, such as a Mercury-like 3:2 spin-orbit resonance, because such rotation would generate tidal heating far in excess of the planet's observed flux. The lack of observed large tidal heating also constrains the planet's eccentricity, albeit with some uncertainty due to the planet's uncertain tidal response. As long as LHS 3844b's tidal dissipation efficiency lies within one order of magnitude of Earth's, its eccentricity has to be smaller than $\sim0.001$.
This means that even if LHS 3844b was in pseudo-synchronous rotation, it has to rotate so slowly as to be effectively 1:1 tidally locked; more likely, given that rocky planets probably cannot be in pseudo-synchronous rotation \citep{makarov2013}, LHS 3844b has to be 1:1 tidally locked.

In addition the Spitzer data provide evidence for either tidal heating or space weathering on LHS 3844b. The secondary eclipse depth only provides a weak constraint, as all models we consider in detail match the secondary eclipse to within $2\sigma$ (Table \ref{tab:chi2 compare}). Instead we compare models directly to the observed phase curve and find that a geologically fresh basalt surface on a circular orbit provides a poor fit to the data. We therefore propose two hypotheses: either LHS 3844b still retains a small-but-nonzero eccentricity which drives weak tidal heating, requiring a third body such as a companion planet, or LHS 3844b's surface has been darkened by space weathering. Current data are insufficient to rule out tidal heating, but we consider space weathering more plausible.

Our results thus suggest that LHS 3844b is a potential exoplanet analog to the Moon and Mercury in our own Solar System, with a similarly darkened and space-weathered surface. Future observations will be able to test and refine this interpretation in a number of ways. Radial velocity measurements can be used to constrain the planet's eccentricity and rule out the presence of planetary companions. Similarly, JWST observations can identify space weathering through its impact on the planet's secondary eclipse spectrum, while tidal heating can be inferred through more precise thermal phase curves. Such measurements should be possible not only for LHS 3844b, but also for other exoplanet bare rock candidates including GJ 1252b \citep{crossfield2022}, TRAPPIST-1b \citep{greene2023}, and TRAPPIST-1c \citep{zieba2023}.

%%If you wish to include an acknowledgments section in your paper,
\acknowledgments
DK was supported by the National Natural Science Foundation of China (NSFC) under grant no. 42250410318. RYH was supported by a grant from the STScI (JWST Program 1846) under NASA contract NAS5-03127. Part of this research was carried out at the Jet Propulsion Laboratory, California Institute of Technology, under a contract with the National Aeronautics and Space Administration (80NM0018D0004).


%% for the paper.  Each keyword is check against the master list during
%% copy editing.  Individual instruments can be provided in parentheses,
%% after the keyword, but they are not verified.

% \vspace{5mm}
% \facilities{HST(STIS), Swift(XRT and UVOT), AAVSO, CTIO:1.3m,

% \software{IRAF, cloudy, IDL}



\appendix

\section{Effect of surface heat capacity}
\label{appendix:heatcapacity}

The calculations presented in the main text assume a low surface heat capacity, $\Gamma= 100$ J s$^{-1/2}$ K$^{-1}$ m$^{-2}$, compatible with a surface composed of regolith. To test this assumption we ran a set of simulations which compare a low heat capacity surface, $\Gamma= 100$ J s$^{-1/2}$ K$^{-1}$ m$^{-2}$,  against a high heat capacity surface, $\Gamma= 2420$ J s$^{-1/2}$ K$^{-1}$ m$^{-2}$, representing solid rock \citep{pierrehumbert2010principles}. To isolate the effect of heat capacity these simulations do not include tidal heating.

\begin{figure}[htbp]
\centering
\includegraphics[width=\textwidth]{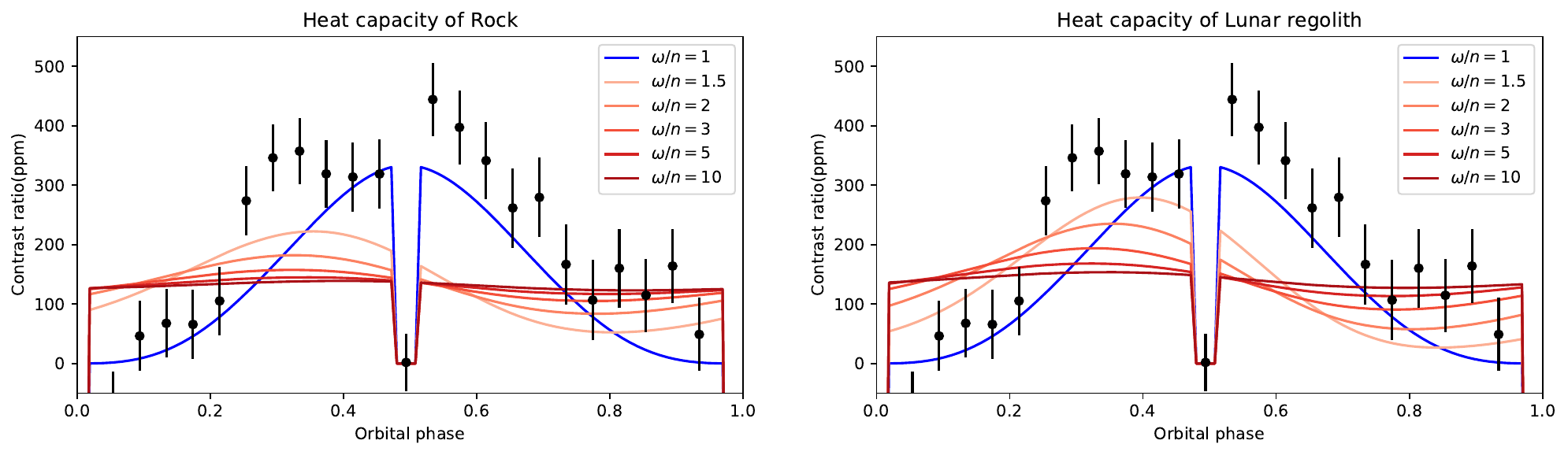}
\caption{Phase curve of LHS3844b for different rotational states and for two different surface heat capacities. Left panel assumes a heat capacity of rock, $\Gamma= 2420$ J s$^{-1/2}$ K$^{-1}$ m$^{-2}$, right panel assumes a heat capacity of regolith $\Gamma= 100$ J s$^{-1/2}$ K$^{-1}$ m$^{-2}$. Tidal heating is disabled. All other parameters are the same as in Figure \ref{fig:bare rock}.}
\label{heat capacity}
\end{figure}

In line with previous work \citep[][]{selsis2013effect}, Figure \ref{heat capacity} shows that the impact of heat capacity becomes noticeable at rapid rotation. The two sets of simulations visibly differ for $\omega/n=3/2$ (i.e., in a 3:2 spin-orbit resonance), with a higher heat capacity leading to a smaller day-night flux variation as well as a larger hot spot shift. These differences become increasingly indistinguishable at extremely rapid rotation, such that at $\omega/n=10$ both phase curves are essentially flat.

Nonetheless, we do not expect the assumed surface heat capacity to strongly affect our results for LHS 3844b. Because our tidal heating calculations rule out spin-orbit resonances and constrain the planet's eccentricity to be less than $\sim 10^{-3}$ (see main text), LHS 3844b's rotation rate in pseudo-synchronous rotation has to be slower than $\omega/n < $1.000006. Based on Figure \ref{heat capacity}, at such low rotation rates the effect of surface heat capacity should be negligible.

\section{Model Validation}
\label{validation}

To validate our bidirectional reflectance calculations, we compare them against a set of calculations performed with the model from \citet{hu2012theoretical} for a hypothetical bare rock exoplanet. The assumed planet and star parameters correspond to a roughly Earth-sized around a cool M-type star, $T_{\mathrm{eff}}=2800 K, M_s=0.1009 M_{\bigodot}, R_s=0.1230 R_{\bigodot}, R_p=0.977 R_{\bigoplus}$.

%$T_{eff}=2800\pm 29K, M_s=0.1009\pm0.0024M_{\bigodot}, R_s=0.1230\pm0.0022R_{\bigodot}, R_p=0.977\pm0.022R_{\bigoplus}$.

\begin{figure}[htbp]
\centering
\includegraphics[width=0.7\textwidth]{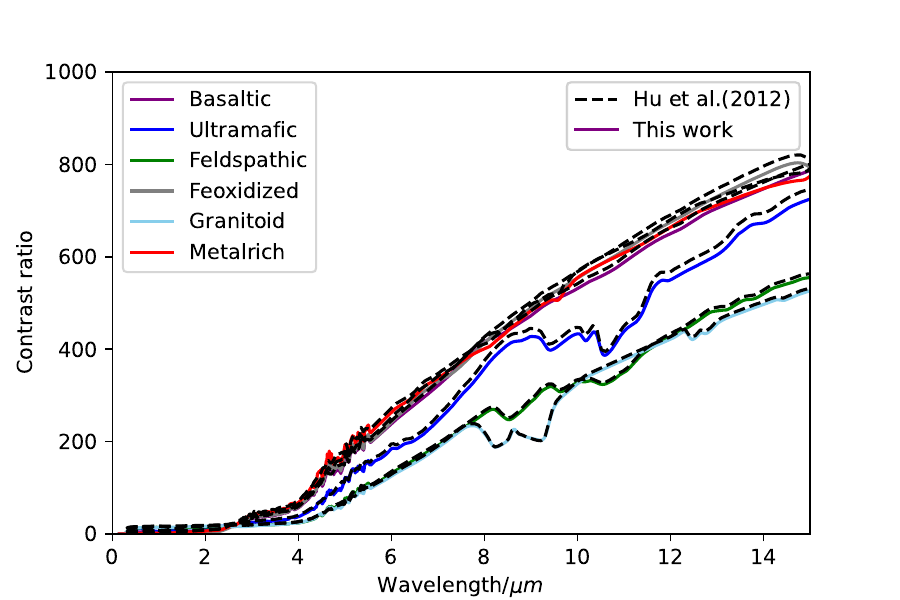}
\caption{Comparison of emission spectra for a hypothetical rocky exoplanet according to our model (colored lines) versus the model of \citet{hu2012theoretical} (black lines).}
\label{compare}
\end{figure}

Dashed black lines in Figure~\ref{compare} show emission spectra computed with the numerical model of \citet{hu2012theoretical}, solid colored lines show our corresponding results. Both sets of calculations use the same stellar spectrum and single scattering albedo for a given surface material. We find that both sets of calculations agree closely on the variation of contrast ratio with respect to wavelength, and moreover agree on the relative flux difference between different surface types.
In terms of model differences, we find that our calculations systematically produce a slightly lower flux than the calculations with \citet{hu2012theoretical}'s model. The maximum flux difference at a single wavelength in the infrared occurs in the wiggles around 5 micron and amounts up to 10\% flux difference, presumably due to the models' different spectral resolutions or spectral averaging. The maximum broadband-averaged flux difference is much smaller, about 2\%. This level of model agreement is sufficient, given that it is much smaller than the typical precision of rocky planet emission spectra with JWST \citep{greene2023,zieba2023}.
%
%In terms of model differences, we find that our calculations systematically produce a slightly lower flux than the calculations with \citet{hu2012theoretical}'s model. In the infrared, the maximum flux difference at a single wavelength between the two models amounts up to 10\%, occurring at the wiggles around 5 microns, while the broadband-averaged flux difference is about 2\%. Overall, the two models thus agree to better than 3\%, which is smaller than the typical error for a rocky planet's emission spectrum with JWST \citep{greene2023,zieba2023}.

To explain the remaining model differences, we hypothesize two plausible sources of error. First, we find that the relative offset between the two models is also robust at short wavelengths (less than 2$\mu m$), where the contrast ratio is dominated by reflected light. A model bias that is independent of wavelength suggests that the two models are using slightly different parameters or constants; for example, rounding off a fundamental constant like the Astronomical Unit, the Solar Radius, or Earth's Radius at the second digit would induce an $\mathcal{O}(1)\%$ error, which in turn is squared and thus could result in up to $10\%$ error in the calculated contrast ratio.
Second, Figure~\ref{compare} also shows some wavelength-dependent model differences. For example, for a Granitoid surface the two models essentially agree at 9 $\mu$m but slightly disagree at 14 $\mu$m, whereas for an Ultramafic surface the offset is roughly the same at both wavelengths.
Unfortunately, we are not able to explain this source of error, as it could be caused by a large number of possible issues such as differences in horizontal (latitude-longitude) resolution, spectral resolution, or (in our model's case) biases in the surface temperature due to the model's finite vertical resolution.

\bibliography{zoterolibrary_latest}
\bibliographystyle{aasjournal}



\end{document}